\documentstyle[12pt]{article}
\begin{document}
\begin{flushright}
hep-th/0510164\\ BHU-SNB/Preprint
\end{flushright}
\vskip 1cm
\begin{center}
{\bf \Large {Superfield approach to  nilpotent symmetries for QED
from a single \\restriction: an alternative to the horizontality
condition}}

\vskip 1.5cm

{\bf R.P.Malik} \\ {\it S. N. Bose National Centre for Basic
Sciences,}
\\ {\it Block-JD, Sector-III, Salt Lake, Calcutta-700 098, India}\\

\vspace{.2cm}

{and}\\

\vspace{.2cm}

 {\it  Centre of Advanced Studies, Physics
Department,}\\ {\it Banaras Hindu University, Varanasi- 221 005,
India}\\ {\bf E-mail address: malik@bhu.ac.in}

\vskip 1.5cm

\end{center}

\noindent {\bf Abstract}: We derive {\it together} the  exact
local, covariant, continuous and off-shell nilpotent
Becchi-Rouet-Stora-Tyutin (BRST) and anti-BRST symmetry
transformations for the $U(1)$ gauge field ($A_\mu$), the
(anti-)ghost fields ($(\bar C)C$) and the Dirac fields ($\psi,
\bar\psi$) of the Lagrangian density of a four ($3 +
1$)-dimensional QED by exploiting {\it a single} restriction on
the six ($4, 2)$-dimensional supermanifold. A set of four even
spacetime coordinates $x^\mu$ ($\mu = 0, 1, 2, 3)$ and two odd
Grassmannian variables $\theta$ and $\bar\theta$ parametrize this
six dimensional supermanifold. The new gauge invariant restriction
on the above supermanifold owes its origin to the (super)
covariant derivatives and their intimate relations with the
(super) 2-form curvatures $(\tilde F^{(2)})F^{(2)}$ constructed
with the help of 1-form (super) gauge connections $(\tilde
A^{(1)})A^{(1)}$ and (super) exterior derivatives $(\tilde d)d$.
The results obtained by exploiting (i) the horizontality
condition, and (ii) one of its consistent extensions, are shown to
be a simple consequence of this new {\it single} restriction on
the above supermanifold. Thus, our present endeavour provides an
alternative to (and, in some sense, generalization of) the
horizontality condition of the usual superfield formalism applied
to the derivation of BRST symmetries.\\

\baselineskip=16pt



\newpage

\noindent
{\bf 1 Introduction}\\

\noindent The usual superfield approach [1-6] to
Becchi-Rouet-Stora-Tyutin (BRST) formalism provides the
geometrical origin and interpretations for the nilpotent
(anti-)BRST symmetry transformations (and their corresponding
generators) for the $p$-form ($p = 1, 2,...)$ gauge fields and
corresponding (anti-)ghost fields of the $p$-form interacting
gauge theories \footnote{Such a class of 1-form gauge theories
(that provide the theoretical basis for the three out of four
fundamental interactions of nature) is endowed with the first
class constraints in the language of Dirac's prescription for
classification scheme [7,8]. These constraints generate the local
gauge symmetries which dictate the interaction term in the theory.
In fact, the interaction term arises due to the coupling of the
$1$-form gauge fields with the conserved Noether currents
constructed by the matter (and other relevant) fields when one
demands the local gauge invariance in the theory.} which include
matter fields as well. This approach, however, does not shed any
light on the nilpotent symmetry transformations associated with
the matter fields, present in the above interacting gauge
theories. It has been a challenging problem to derive them
cogently within the framework of the superfield approach to BRST
formalism.

The above usual superfield formalism has been exploited
extensively for the derivation of the nilpotent (anti-)BRST
symmetries in the context of four $(3 + 1)$-dimensional (4D)
1-form and 2-form (non-)Abelian gauge theories which are, in
general, considered on the six $(4, 2)$-dimensional (6D)
supermanifold [1-6]. The latter is parametrized by the superspace
variables $Z^M = (x^\mu, \theta, \bar\theta)$ where the four even
(bosonic) variables $x^\mu\; (\mu = 0, 1, 2, 3)$ correspond to the
4D spacetime variables and two odd Grassmannian ($\theta^2 =
\bar\theta^2 = 0, \theta \bar\theta + \bar\theta \theta = 0$)
variables are the additional coordinates on the supermanifold. The
nilpotent (anti-)BRST symmetries for the 1-form 4D non-Abelian
gauge fields and the corresponding (anti-)ghost fields emerge from
the horizontality condition [1-6] on the 6D supermanifold which
enforces the equality ($\tilde F^{(2)} = F^{(2)}$) of the 2-form
super curvature $\tilde F^{(2)} = \tilde d \tilde A^{(1)} + \tilde
A^{(1)} \wedge \tilde A^{(1)}$  (constructed with the help of the
super exterior derivative $\tilde d$ and the 1-form super
connection $\tilde A^{(1)}$) to the ordinary 2-form curvature
$F^{(2)} = d A^{(1)} + A^{(1)} \wedge A^{(1)}$ (constructed with
the help of the ordinary exterior derivative $d$ and the 1-form
connection $A^{(1)}$). The above arguments (with the theoretical
arsenal of the horizontality condition) have also been applied to
the case of 2-form Abelian gauge theory in a straightforward
manner (see, e.g., [6] for details).

For the discussion of any arbitrary four dimensional $p$-form
[$A^{(p)} = \frac{1}{p!} (dx^{\mu_{1}} \wedge dx^{\mu_{2}}
....\wedge dx^{\mu_{p}})\; A_{\mu_{1} \mu_{2} .....\mu_{p}}$]
Abelian gauge theory, within the framework of the usual superfield
approach to BRST symmetries, one constructs a $(p + 1)$-form super
curvature $\tilde F^{(p + 1)} = \tilde d \tilde A^{(p)}$ with the
help of a super exterior derivative $\tilde d$ and the super 6D
$p$-form connection $\tilde A^{(p)}$ on the 6D supermanifold. This
is subsequently equated, due to the so-called horizontality
condition [1-6], to the ordinary four dimensional $(p + 1)$ form
curvature $F^{(p + 1)} = d A^{(p)}$ constructed with the help of
the ordinary exterior derivative $d = dx^\mu \partial_\mu$ (with
$d^2 = 0$) and the ordinary 4D $p$-form connection $A^{(p)}$. The
covariant reduction of the 6D super curvature to the ordinary 4D
curvature, through the equality $\tilde F^{(p + 1)} = F^{(p + 1)}$
due to the horizontality condition, leads to the derivation of the
nilpotent (anti-)BRST symmetry transformations for the $p$-form
Abelian gauge field and the corresponding (anti)commuting
(anti-)ghost fields of the given $p$-form 4D Abelian gauge theory.

The horizontality condition of the above superfield approach has
been christened as the soul-flatness condition in [9] which
amounts to setting equal to zero the Grassmannian components of
the (anti)symmetric super curvature tensor that constitutes the
super 2-form $\tilde F^{(2)}$ (corresponding to a given 1-form
gauge theory). The covariant reduction of $\tilde F^{(2)}$
(defined on the 6D supermanifold) to the ordinary 2-form curvature
$F^{(2)}$ (defined on the 4D ordinary spacetime manifold) leads to
the geometrical origin and interpretations for (i) the internal
nilpotent (anti-)BRST symmetry transformations for the 4D ordinary
fields as the translations of the corresponding 6D superfields
along the Grassmannian directions of the 6D supermanifold, (ii)
the nilpotent (anti-)BRST charges as the translation generators
along $\theta$ and $\bar\theta$ directions of the 6D
supermanifold, (iii) the nilpotency property as a couple of
successive translations along a particular Grassmannian direction
of the supermanifold, and (iv) the anticommutativity property of
the (anti-)BRST symmetries (and their generators) as the
anticommutativity encoded in the translational generators along
the $\theta$ and $\bar\theta$ directions of the supermanifold (cf.
(4.24) below). These beautiful connections between the geometrical
objects on the 6D supermanifold and some key properties associated
with the internal nilpotent symmetry transformations of the BRST
formalism in the ordinary 4D spacetime are, however, confined {\it
only} to the gauge fields and the (anti-)ghost fields of the
theory within the framework of the {\it usual} superfield approach
to BRST formalism [1-6].

In a very recent set of papers [10-14], the usual superfield
formalism has been generalized to the {\it augmented} superfield
formalism \footnote{Any mathematically consistent generalization
of the usual superfield approach to BRST formalism.} where
additional restrictions on the 6D supermanifolds have been invoked
which have been found to be consistent with (and complementary to)
the horizontality condition. This augmented version of the
superfield approach enables one to  derive the nilpotent
(anti-)BRST symmetry transformations for {\it all} the fields of
the (non-)Abelian gauge theories [10-14] (as well as the
reparametrization invariant (supersymmetric) theories [11]) while
keeping the geometrical interpretations of the (anti-)BRST
symmetries (and their generators) intact. These additional
restrictions on the 6D supermanifold owe their origin to the
equality of (i) the conserved and gauge invariant matter currents
[10] (and other conserved quantities [11]), and (ii) the gauge
(i.e. BRST) invariant quantities constructed with the help of the
(super) covariant derivatives [12-14]. The former restrictions
allow a logically consistent derivation of the (anti-)BRST
symmetry transformations for the matter fields whereas the latter
lead to the derivations that are mathematically unique. Both the
above extensions have their own merits and advantages.

The purpose of the present paper is to derive the off-shell
nilpotent and anticommuting (anti-)BRST symmetry transformations
for {\it all} the fields of the 4D QED (that includes Dirac fields
as matter fields) from  a {\it single} gauge (i.e. BRST) invariant
restriction on the 6D supermanifold. We obtain all the nilpotent
symmetry transformations that are derived by exploiting (i) the
horizontality condition, and (ii) one of its consistent
generalizations [10], separately. In fact, the consequences of
both the above independent restrictions (i.e. (i) and (ii)) emerge
very naturally from our present single restriction (cf. (4.1)
below). Our present investigation is essential primarily on four
counts. First and foremost, the horizontality condition, as
discussed earlier, does not shed any light on the derivation of
the nilpotent symmetry transformations associated with the matter
fields of a given interacting gauge theory whereas our present
single restriction on the 6D supermanifold does precisely that.
Second, the single restriction (cf. (4.1) below) imposed on the 6D
supermanifold is a gauge (i.e. BRST) invariant condition that is
more physical than the horizontality condition which happens to be
{\it intrinsically} a gauge covariant restriction. Third, our
present single restriction is a nice simplification of our
previous attempts [10-14] where two separate restrictions were
imposed on the supermanifold for the derivation of all the
nilpotent transformations in the context of (non-)Abelian gauge
(and reparametrization invariant) theories. Finally, the
horizontality condition and one of its consistent extensions [10]
are, in some sense, unified together in our present single
restriction. Thus, the imposition of our present single
restriction (cf. (4.1) below) on the 6D supermanifold is
aesthetically and physically more appealing than the imposition of
the  horizontality condition alone.

The contents of our present paper are organized as follows. In section 2,
we set up the notations and conventions by recapitulating the bare essentials
of the (anti-)BRST symmetry transformations in the framework
of Lagrangian formulation for QED with Dirac fields. Section 3
is devoted to the definition of suitable superfields and their expansions, in
terms of  the basic and some secondary fields, along the Grassmannian
directions of the supermanifold. The central results of our investigation are
contained in section 4 where we derive the nilpotent (anti-)BRST
transformations for {\it all} the fields of the above QED from a single
restriction (cf. (4.1) below) on the supermanifold. Finally, we
make some concluding remarks and point
out a few future directions for further investigations in section 5.\\

\noindent {\bf 2 (Anti-)BRST symmetries in Lagrangian formulation:
a brief sketch}\\

\noindent Let us begin with the (anti-)BRST invariant Lagrangian
density ${\cal L}_{B}$ for the {\it interacting} four ($3 +
1)$-dimensional (4D) $U(1)$ gauge theory (QED) in the Feynman
gauge \footnote{ We adopt here the notations and conventions such
that the flat Monkowskian metric $\eta_{\mu\nu} = $ diag $(+1, -1,
-1, -1)$ for the 4D spacetime manifold and $F_{0i} = \partial_0
A_i - \partial_i A_0 = E_i, F_{ij} = \epsilon_{ijk} B_k, B_i =
\frac{1}{2} \epsilon_{ijk} F_{jk}$ are the electric ($E_i$) and
magnetic ($B_i$) components of the field strength tensor
$F_{\mu\nu}$. Here $\epsilon_{ijk}$ is the totally antisymmetric
Levi-Civita tensor (with $\epsilon_{123} = + 1$) on the 3D
subspace of the 4D Minkowskian space. Furthermore, the Greek
indices $\mu, \nu, ....= 0, 1, 2, 3$, present in (2.1), stand for
the spacetime directions and Latin indices $i, j, k...= 1, 2,3$
correspond only to the space directions on the 4D spacetime
manifold.} [9,15,16] $$
\begin{array}{lcl}
{\cal L}_{B} = - \frac{1}{4}\; F^{\mu\nu} F_{\mu\nu} + \bar \psi
\;(i \gamma^\mu D_\mu - m)\; \psi + B \;(\partial \cdot A) +
\frac{1}{2}\; B^2 - i \;\partial_{\mu} \bar C \partial^\mu C
\end{array} \eqno(2.1)
$$ where $D_\mu \psi = \partial_\mu \psi + i e A_\mu \psi$ is the
covariant derivative on the Dirac field $\psi (x)$ with charge $e$
and mass $m$. The $U(1)$ gauge field $A_\mu$ couples to the matter
conserved current $J_\mu = \bar \psi \gamma_\mu \psi$ (constructed
by the Dirac fields $(\psi, \bar\psi)$) with the coupling strength
$e$. This  coupling generates an interaction term $- e \bar\psi
\gamma^\mu A_\mu \psi$ in the theory which exists basically due to
the requirement of the local $U(1)$ gauge invariance. The
$\gamma$'s in (2.1) are the usual Dirac $4 \times 4$ matrices. The
2-form $F^{(2)} = d A^{(1)} = \frac{1}{2} (dx^\mu \wedge dx^\nu)\;
F_{\mu\nu}$, constructed with the help of the exterior derivative
$d = dx^\mu \partial_\mu$ (with $d^2 = 0$) and 1-form $A^{(1)} =
dx^\mu A_\mu$, defines the field strength tensor $F_{\mu\nu} =
\partial_\mu A_\nu - \partial_\nu A_\mu$ for the $U(1)$ gauge
field $A_\mu$. The Nakanishi-Lautrup auxiliary field $B$
linearizes the gauge-fixing term $- \frac{1}{2} (\partial \cdot
A)^2$ of the Lagrangian density (2.1) and the fermionic (i.e. $
C^2 = \bar C^2 = 0, C \bar C + \bar C C = 0$) (anti-)ghost fields
$(\bar C)C$ are required to maintain the unitarity and ``quantum''
gauge (i.e. BRST) invariance together, for a given physical
process, at any arbitrary order of perturbative computation
\footnote{ The importance of the (anti-)ghost fields emerges in
its full blaze of glory in the context of perturbative
computations, connected with a given physical process,
 that is allowed by the {\it interacting} non-Abelian gauge theory.
In fact, for the proof of unitarity of such a kind of physical process, one
requires a Feynman loop diagram constructed by purely the fermionic
(anti-)ghost fields corresponding to
each such loop diagram existing in the theory due to a purely
bosonic non-Abelian gauge (gluon) field (see, e.g., [17] for details).}.

The above Lagrangian density (2.1) for QED with Dirac fields,
respects the following
infinitesimal, off-shell
nilpotent ($s_{(a)b}^2 = 0$), anticommuting ($s_b s_{ab} + s_{ab} s_b = 0$),
local, continuous and covariant (anti-)BRST ($s_{(a)b}$)
symmetry transformations
\footnote{We follow here the notations adopted in [15,16]. In fact, the BRST
prescription is to replace the local gauge parameter of the original gauge
theory by an anticommuting ($ \eta C + C \eta = 0, \eta \psi + \psi \eta = 0,$
etc.) spacetime independent parameter $\eta$ and the (anti-)ghost fields.
Thus, in its totality, the (anti-)BRST transformations $\delta_{(A)B}$ are a
product ($\delta_{(A)B} = \eta s_{(a)b}$)
of $\eta$ and the nilpotent $s_{(a)b}^2 = 0$ transformations $s_{(a)b}$.}
[9,15,16]
$$
\begin{array}{lcl}
s_{b} A_{\mu} &=& \partial_{\mu} C\; \qquad s_{b} C = 0\; \qquad
s_{b} \bar C = i B\;  \qquad s_b \psi = - i e C \psi \nonumber\\
s_b \bar \psi &=& - i e \bar \psi C\; \qquad   s_{b} B = 0\; \quad
\;s_{b} F_{\mu\nu} = 0\; \quad s_b (\partial \cdot A) = \Box C
\nonumber\\ s_{ab} A_{\mu} &=& \partial_{\mu} \bar C\; \qquad
s_{ab} \bar C = 0\; \qquad s_{ab} C = - i B\;  \qquad s_{ab} \psi
= - i e \bar C \psi \nonumber\\ s_{ab} \bar \psi &=& - i e \bar
\psi \bar C\; \qquad  s_{ab} B = 0\; \quad \;s_{ab} F_{\mu\nu} =
0\; \quad s_{ab} (\partial \cdot A) = \Box \bar C
\end{array}\eqno(2.2)
$$
because it transforms to a total derivative. Some noteworthy points, at this
juncture, are in order now. First, under the nilpotent
(anti-)BRST transformations,
the kinetic energy term of the (non-)Abelian gauge fields remains invariant.
More precisely, for the Abelian gauge theory, it is the field strength tensor
$F_{\mu\nu}$ itself that remains unchanged. Second, the gauge-fixing term
$(\partial \cdot A)$, on the other hand,
transforms under the (anti-)BRST transformations.
Finally, the cohomological
operator $d = dx^\mu \partial_\mu$ (with $d^2 = 0$) and the nilpotent
($s_{(a)b}^2 = 0$) (anti-)BRST transformations $s_{(a)b}$ are inter-connected.
This is due to the fact that the electric $E_i$ and magnetic $B_i$ fields
(which are components of $F_{\mu\nu}$)
remain invariant under the  transformations $s_{(a)b}$ and they owe their
origin, primarily, to the nilpotent ($d^2 = 0$) cohomological operator $d$
because $F^{(2)} = d A^{(1)} = \frac{1}{2} (dx^\mu \wedge dx^\nu) F_{\mu\nu}$.

According to the Noether's theorem, the above continuous symmetry
transformations lead to the derivation of the conserved
(anti-)BRST charges $Q_{(a)b}$ which are found to be nilpotent
($Q_{(a)b}^2 = 0$) of order two. These charges generate the above
continuous nilpotent transformations. For a generic local field
$\Omega  = A_\mu, C, \bar C, \psi, \bar \psi, B$ of the Lagrangian
density (2.1), the infinitesimal transformations (2.2) can be
expressed in terms of $Q_{(a)b}$, as
$$
\begin{array}{lcl}
s_{r}\; \Omega (x) = - i\;\Bigl [\; \Omega (x), Q_{r}\; \Bigr
]_{\pm}\;\; \qquad \;\;\;r = b, ab
\end{array} \eqno(2.3)
$$
where the subscripts $(+)-$, on the square bracket, stand for this
bracket to correspond to
an (anti)commutator for the generic local field $\Omega (x)$
of the Lagrangian density (2.1) being
(fermionic)bosonic in nature. For our present discussions, the exact
expressions for the conserved, nilpotent
and anticommuting ($Q_b Q_{ab} + Q_{ab} Q_b = 0$) (anti-)BRST charges
$Q_{(a)b}$ are not important but their explicit
forms can be found in [9,15,16].\\

\noindent {\bf 3 From ordinary basic fields to superfields: super
expansion}\\

\noindent To derive the above local, continuous, covariant,
nilpotent and anticommuting (anti-)BRST infinitesimal
transformations (2.2) within the framework of superfield
formalism, first of all, we generalize the basic local fields
$A_\mu, C, \bar C, \psi, \bar \psi$ of the Lagrangian density
(2.1), defined on the 4D spacetime manifold, to the corresponding
superfields ${\cal B}_\mu, {\cal F}, \bar {\cal F}, \Psi, \bar
\Psi$ defined on the six dimensional supermanifold parametrized by
the superspace variables $Z^M = (x^\mu, \theta, \bar\theta)$.
These superfields can be expanded in terms of the basic fields
($A_\mu, C, \bar C, \psi, \bar\psi)$ and extra secondary fields,
along the Grassmannian directions, as follows [4,3,10] $$
\begin{array}{lcl}
{\cal B}_{\mu} (x, \theta, \bar \theta) &=& A_{\mu} (x) + \theta\;
\bar R_{\mu} (x) + \bar \theta\; R_{\mu} (x) + i \;\theta \;\bar
\theta S_{\mu} (x) \nonumber\\ {\cal F} (x, \theta, \bar \theta)
&=& C (x) + i\; \theta \bar B_1 (x) + i \;\bar \theta\; B_2 (x) +
i\; \theta\; \bar \theta \;s (x) \nonumber\\ \bar {\cal F} (x,
\theta, \bar \theta) &=& \bar C (x) + i \;\theta\;\bar B_2 (x) +
i\; \bar \theta \;B_1 (x) + i \;\theta \;\bar \theta \;\bar s (x)
\nonumber\\
 \Psi (x, \theta, \bar\theta) &=& \psi (x)
+ i \;\theta\; \bar b_1 (x) + i \;\bar \theta \; b_2 (x) + i
\;\theta \;\bar \theta \;f (x) \nonumber\\ \bar \Psi (x, \theta,
\bar\theta) &=& \bar \psi (x) + i\; \theta \;\bar b_2 (x) + i
\;\bar \theta \; b_1 (x) + i\; \theta \;\bar \theta \;\bar f (x).
\end{array} \eqno(3.1)
$$
It is obvious that the superfield ${\cal B}_\mu (x,\theta, \bar\theta)$
is bosonic
and the rest of the above superfields are fermionic
(i.e. ${\cal F}^2 = \bar {\cal F}^2 = \Psi^2 = \bar \Psi^2 =0$) in nature.

A few salient features of the above expansions are as follows:

\noindent
(i) In the limit $(\theta, \bar\theta) \to 0$, we do retrieve the basic
fields of the Lagrangian density (2.1) that are defined on the 4D ordinary
Minkowskian spacetime manifold.

\noindent
(ii) In the above expansion, the total number of fermionic fields
$(\psi, \bar\psi, f, \bar f, C, \bar C, s, \bar s, R_\mu, \bar R_\mu)$
do match with the bosonic fields $(b_1, \bar b_1, b_2, \bar b_2,
B_1, \bar B_1, B_2, \bar B_2, A_\mu, S_\mu$).

\noindent
(iii) The above straightforward observation in (ii) is an essential requirement
for any arbitrary superfield theory to be discussed in the framework
of supersymmetry.

It is important to generalize the exterior derivative $d = dx^\mu
\partial_\mu$ and the 1-form connection $A^{(1)} = dx^\mu A_\mu$,
defined on the ordinary 4D Minkowskian flat manifold, to the six
$(4,  2)$-dimensional supermanifold. This is required because, as
discussed and emphasized earlier after (2.2), the above
geometrical quantities have relevance with the (anti-)BRST
symmetry transformations. Thus, these quantities on the above
supermanifold, are $$
\begin{array}{lcl}
\tilde d &=& \;d Z^M \;\partial_{M} = d x^\mu\; \partial_\mu\; +
\;d \theta \;\partial_{\theta}\; + \;d \bar \theta
\;\partial_{\bar \theta} \nonumber\\ \tilde A^{(1)} &=& d Z^M\;
\tilde A_{M} = d x^\mu \;{\cal B}_{\mu} (x , \theta, \bar \theta)
+ d \theta\; \bar {\cal F} (x, \theta, \bar \theta) + d \bar
\theta\; {\cal F} ( x, \theta, \bar \theta)
\end{array}\eqno(3.2)
$$
which reduce to $d = dx^\mu \partial_\mu$ and $A^{(1)} = dx^\mu A_\mu$
in the limit $(\theta, \bar\theta) \to 0$. It is clear, therefore, that (i)
$\tilde d$ and $\tilde A^{(1)}$ are a set of consistent 6D
superspace generalization
of the 4D quantities $d $ and $A^{(1)}$
defined on the ordinary space, and (ii) the superspace
derivative $\partial_M$ and supervector superfield $\tilde A_M$
have the component
multiplets $(\partial_\mu, \partial_\theta, \partial_{\bar\theta})$
and $({\cal B}_\mu, {\cal F}, \bar {\cal F})$, respectively. Both
the quantities, defined in (3.2), would be useful in the next section.\\

\noindent {\bf 4 Gauge invariant restriction on supermanifold:
nilpotent symmetries}\\

\noindent To provide the geometrical interpretation for the
nilpotent symmetry transformations (2.2) within the framework of
the superfield approach to BRST formalism, we have to exploit a
certain specific restriction on the supermanifold. To this end in
mind, we begin with the following gauge invariant restriction on
the six $(4, 2)$-dimensional supermanifold: $$
\begin{array}{lcl}
\bar \Psi (x, \theta, \bar \theta) \tilde D \; \tilde D \; \Psi
(x, \theta, \bar\theta) = \bar \psi (x) \; D \; D \; \psi (x)
\end{array}\eqno(4.1)
$$
where the (super) covariant derivatives on the six-dimensional
supermanifold (i.e. $\tilde D$) and
ordinary 4D Minkowskian spacetime manifold (i.e. $D$), are
$$
\begin{array}{lcl}
\tilde D = \tilde d + i\; e \; \tilde A^{(1)} (x, \theta,
\bar\theta)\; \qquad D = d + i\; e\; A^{(1)} (x).
\end{array}\eqno(4.2)
$$ In the above, the symbols $\tilde d$ and $\tilde A^{(1)}$ are
defined in (3.2) on the $(4,  2)$-dimensional supermanifold and
corresponding 4D quantities are: $d = dx^\mu \partial_\mu, A^{(1)}
= dx^\mu A_\mu$ .

It is obvious that
the r.h.s. of (4.1) is a $U(1)$ gauge invariant quantity which can be
explicitly expressed, in terms of the 2-form curvature
$F^{(2)} = d A^{(1)}$, as
$$
\begin{array}{lcl}
\bar \psi (x) \; D \; D \; \psi (x) = \frac{1}{2}\;i e\;
(dx^\mu \wedge dx^\nu)\;
\bar \psi (x)\; F_{\mu\nu} (x)\; \psi (x) \equiv
i e\; \bar \psi\; F^{(2)}\; \psi.
\end{array}\eqno(4.3)
$$ It will be noted that (i) the r.h.s. of (4.3) possesses only
the 2-form differentials $\frac{1}{2} (dx^\mu \wedge dx^\nu)$ in
terms of spacetime variables, and (ii) the well-known relation $D
D \psi = i e F^{(2)} \psi$ has been used in the above derivation.
In contrast, the l.h.s will lead to the 2-form differentials
$\frac{1}{2} (dZ^M \wedge dZ^N)$ which will contain all the
possible combinations of 2-forms, constructed by superspace
differentials (i.e. $dx^\mu \wedge dx^\nu, dx^\mu \wedge d\theta,
dx^\mu \wedge d\bar\theta, d \bar\theta \wedge d \bar\theta, d
\theta \wedge d\bar\theta, d \theta \wedge d\theta$).

The explicit form of the l.h.s., in terms of the
component multiplet superfields ${\cal B}_\mu, {\cal F}, \bar {\cal F}$,
the superspace differentials and the partial
derivatives $\partial_\mu, \partial_\theta, \partial_{\bar\theta}$,
can be written as
$$
\begin{array}{lcl}
&& \bar \Psi (dx^\mu) (\partial_\mu + i e {\cal B}_\mu) \wedge
\Bigl [ dx^\nu (\partial_\nu + i e {\cal B}_\nu) \Psi
+ d \theta (\partial_\theta + i e \bar {\cal F}) \Psi
+ d \bar \theta (\partial_{\bar\theta} + i e {\cal F}) \Psi \Bigr]
\nonumber\\
&+&  \bar \Psi (d\theta) (\partial_\theta + i e \bar {\cal F}) \wedge
\Bigl [ dx^\mu (\partial_\mu + i e {\cal B}_\mu) \Psi
+ d \theta (\partial_\theta + i e \bar {\cal F}) \Psi
+ d \bar \theta (\partial_{\bar\theta} + i e {\cal F}) \Psi \Bigr]
\nonumber\\
&+&  \bar \Psi (d \bar\theta) (\partial_{\bar\theta} + i e {\cal F}) \wedge
\Bigl [ dx^\mu (\partial_\mu + i e {\cal B}_\mu) \Psi
+ d \theta (\partial_\theta + i e \bar {\cal F}) \Psi
+ d \bar \theta (\partial_{\bar\theta} + i e {\cal F}) \Psi \Bigr].
\end{array}\eqno(4.4)
$$
The expansion of the above equation would lead to the coefficients
of $(dZ^M \wedge dZ^N)$ where the superspace variable
$Z^M = (x^\mu, \theta, \bar\theta)$. It is straightforward to note that
the 2-form, constructed {\it only} with the spacetime differentials
$(dx^\mu \wedge dx^\nu)$, would match with
the similar kind of 2-form emerging from the r.h.s. (cf. (4.3)).
The rest of the components of the super 2-form (with the Grassmannian
differentials) will be set equal to zero due to (4.1).

For algebraic convenience, it is useful to first collect the
coefficients of ($d\theta \wedge d\theta$) which can be
succinctly expressed as follows:
$$
\begin{array}{lcl}
- i e \; (d \theta \wedge d\theta)\; \Bigl [\; i \bar \psi \bar
B_2 \psi + \theta\; (L_1) + \bar \theta\; (M_1) - \theta
\bar\theta\; (N_1)\; \Bigr ]
\end{array}\eqno(4.5)
$$
where the explicit expressions for $L_1, M_1$ and $N_1$ are
$$
\begin{array}{lcl}
L_1 = \bar \psi \bar  B_2 \bar b_1 - \bar b_2 \bar B_2 \psi\;
\qquad M_1 = \bar \psi \bar B_2 b_2 - b_1 \bar B_2 \psi - i \bar
\psi \bar s \psi\; \nonumber\\ N_1 = \bar \psi \bar B_2 f + \bar
\psi \bar s \bar b_1 + \bar f \bar B_2 \psi + \bar b_2 \bar s \psi
+ i \bar b_2 \bar B_2 b_2 - i b_1 \bar B_2 \bar b_1.
\end{array}\eqno(4.6)
$$
In the above computation, we have exploited the following inputs:
$$
\begin{array}{lcl}
\partial_\theta \partial_{\bar\theta} \Psi = - i f\; \qquad
\partial_\theta \Psi = i \bar b_1 + i \bar\theta f\; \qquad
\partial_{\bar\theta} \Psi = i b_2 - i \theta f\; \qquad
\partial_\theta {\cal F} = i \bar B_1 + i \bar\theta s.
\end{array}\eqno(4.7)
$$
Ultimately, we have to set equal to zero, separately
and independently, the coefficients
of $(d \theta \wedge d\theta),
[(d \theta \wedge d\theta) (\theta)],
[(d \theta \wedge d\theta) (\bar\theta)]$ and
$[ (d \theta \wedge d\theta) (\theta\bar\theta) ]$.
Restrictions emerging from these
conditions, for $e \neq 0, \psi \neq 0, \bar \psi \neq 0$, are
$$
\begin{array}{lcl}
\bar B_2 = 0\;\; \qquad \;\;\bar s = 0.
\end{array}\eqno(4.8)
$$
The above values, substituted in (3.1), imply that the fermionic superfield
$\bar {\cal F} (x, \theta, \bar\theta)$
becomes an {\it anti-chiral} superfield because
it is constrained to be independent of $\theta$.

In an exactly similar fashion, we can compute the coefficients
of ($d\bar\theta \wedge d\bar\theta$). These are concisely
expressed by the following equation
$$
\begin{array}{lcl}
+ i e \; (d \bar\theta \wedge d\bar\theta)\; \Bigl [\; - i \bar
\psi B_2 \psi + \theta\; (L_2) + \bar \theta\; (M_2) + \theta
\bar\theta\; (N_2)\; \Bigr ]
\end{array}\eqno(4.9)
$$
where the detailed expressions for $L_2, M_2, N_2$, are
$$
\begin{array}{lcl}
L_2 = - \bar \psi B_2 \bar b_1 + \bar b_2 B_2 \psi\; \qquad M_2 =
- \bar \psi  B_2 b_2 + b_1  B_2 \psi + i \bar \psi  s \psi\;
\nonumber\\ N_2 = \bar \psi  B_2 f + \bar \psi  s \bar b_1 + \bar
f B_2 \psi + \bar b_2  s \psi + i \bar b_2  B_2 b_2 + i b_1  B_2
\bar b_1.
\end{array}\eqno(4.10)
$$ To retain the restriction imposed on supermanifold in (4.1), it
is straightforward to note that the coefficients of $(d\bar\theta
\wedge d\bar\theta), [(d\bar\theta \wedge d\bar\theta) (\theta)],
[(d\bar\theta \wedge d\bar\theta) (\bar\theta)]$ and
$[(d\bar\theta \wedge d\bar\theta) (\theta\bar\theta)]$ in (4.9)
would be equal to zero separately and independently. These finally
imply (for $e \neq 0, \psi \neq 0, \bar \psi \neq 0$) $$
\begin{array}{lcl}
B_2 = 0\;\; \qquad \;\; s = 0.
\end{array}\eqno(4.11)
$$
This entails upon the fermionic superfield
${\cal F} (x,\theta,\bar\theta)$ to become {\it chiral} in nature.
Results of (4.8) and (4.11) lead to the following expansions for
the fermionic superfields in (3.1):
$$
\begin{array}{lcl}
{\cal F}^{(c)} (x, \theta) = C (x) + i \theta \bar B_1 (x)\;
\qquad \;\bar {\cal F}^{(ac)} (x, \bar\theta) = \bar C (x) + i
\bar \theta B_1 (x).
\end{array}\eqno(4.12)
$$
The above expansions will be used in our further computations.

Now we focus on the computations of the coefficients of
$(d\theta \wedge d\bar\theta)$. These are expressed in terms of
the fermionic superfield expansion of (4.12), as
$$
\begin{array}{lcl}
- (d\theta \wedge d\bar\theta)\; \Bigl [\;
\bar \Psi \;
\Bigl \{ \bigl (\partial_\theta + i e \bar {\cal F}^{(ac)})
\bigl (\partial_{\bar \theta} + i e  {\cal F}^{(c)})
+
\bigl (\partial_{\bar \theta} + i e  {\cal F}^{(c)} \bigr )
\bigl (\partial_{\theta} + i e  \bar {\cal F}^{(ac)} \bigr )\;
\Bigr \}\;  \Psi\; \Bigr ].
\end{array}\eqno(4.13)
$$
The above equation can be simplified to finally yield
$$
\begin{array}{lcl}
- i e (d \theta \wedge d\bar\theta)\; \Bigl [\; \bar \Psi \; \bigl
(\partial_\theta {\cal F}^{(c)} + \partial_{\bar\theta} \bar {\cal
F}^{(ac)} \bigr )\;\Psi \Bigr ] = 0
\end{array}\eqno(4.14)
$$
where we have used:
 $\partial_\theta \partial_{\bar\theta}
+ \partial_{\bar\theta} \partial_\theta  = 0,
C \bar C = - \bar C C, {\cal F}^{(c)} \bar {\cal F}^{(ac)}
= - \bar {\cal F}^{(ac)} {\cal F}^{(c)}$.
This condition
can be satisfied if and only if $B_1 (x) + \bar B_1 (x) = 0$.
We are free to choose $B_1 (x) = - \bar B_1 (x) = B (x)$
of the Lagrangian density (2.1). Thus, we have
(for $\Psi \neq 0, \bar \Psi \neq 0, e \neq 0$)
$$
\begin{array}{lcl}
&& {\cal F}^{(c)} = C (x) - i\; \theta\;  B (x) \equiv C (x) +
\theta\; (s_{ab} C (x)) \nonumber\\ &&\bar {\cal F}^{(ac)} = \bar
C (x) + i\; \bar \theta \; B (x) \equiv \bar C (x) + \bar\theta\;
(s_b \bar C (x))
\end{array}\eqno(4.15)
$$
which lead to the derivation of the (anti-)BRST transformations (2.2) for
the (anti-)ghost fields $(\bar C)C$ in the framework of the superfield
formalism with restriction (4.1).

We collect the coefficients of $(dx^\mu \wedge d\theta)$ and
$(dx^\mu \wedge d \bar\theta)$ from (4.4) and set them equal to zero
to maintain the consistency with the restriction (4.1). These imply
$$
\begin{array}{lcl}
i e (dx^\mu \wedge d\theta) \bigl [ \bar \Psi \bigl
(\partial_\theta {\cal B}_\mu - \partial_\mu \bar {\cal F}^{(ac)}
\bigr ) \Psi \bigr ] = 0\; \;\;\; \;\;i e (dx^\mu \wedge d\bar
\theta) \bigl [ \bar \Psi \bigl (\partial_{\bar \theta} {\cal
B}_\mu -
\partial_\mu  {\cal F}^{(c)} \bigr ) \Psi \bigr ] = 0.
\end{array}\eqno(4.16)
$$
The above requirements, using the expansions (3.1) and (4.15), lead to
$$
\begin{array}{lcl}
R_{\mu} \;(x) = \partial_{\mu}\; C(x)\; \qquad \;\bar R_{\mu}\;
(x) =
\partial_{\mu}\; \bar C (x)\; \qquad \;S_{\mu}\; (x) = \partial_{\mu}
B\; (x).
\end{array} \eqno(4.17)
$$
Substitution of these values in (3.1) leads to the derivation of
(anti-)BRST symmetry transformations (2.2) for the $U(1)$ gauge field $A_\mu$,
as the superfield ${\cal B}_\mu \to {\cal B}_\mu^{(g)}$. That is:
$$
\begin{array}{lcl}
{\cal B}^{(g)}_{\mu}\; (x, \theta, \bar \theta) = A_{\mu} (x)
+ \;\theta\; (s_{ab} A_{\mu} (x))
+ \;\bar \theta\; (s_{b} A_{\mu} (x))
+ \;\theta \;\bar \theta \;(s_{b} s_{ab} A_{\mu} (x)).
\end{array}\eqno(4.18)
$$
It is worthwhile to emphasize that (i) unlike the fermionic superfields
$(\bar {\cal F}, {\cal F})$ which reduce to (anti-)chiral superfields
after the application of the restriction (4.1), the bosonic superfield
${\cal B}_\mu$ retains its general form (i.e. ${\cal B}_\mu \to
{\cal B}_\mu^{(g)}$) even after application of (4.1), and (ii)
the expansions in (4.15) and (4.18)
have been obtained in earlier works [4,3,10] by exploiting the
horizontality condition
\footnote{In the horizontality condition $\tilde F^{(2)} = F^{(2)}$,
the super 2-form curvature (i.e. $\tilde F^{(2)}
= \tilde d \tilde A^{(1)} =  \frac{1}{2} (d Z^M \wedge d Z^N)
\tilde F_{MN}$) and  the ordinary 2-form curvature (i.e. $F^{(2)} =
d A^{(1)} = \frac{1}{2} (dx^\mu \wedge dx^\nu) F_{\mu\nu}$), are
equated on the supermanifold where $\tilde d$ and $\tilde A^{(1)}$
are defined in (3.2). This restriction implies
$R_{\mu} = \partial_{\mu} C, \bar R_{\mu} = \partial_{\mu}
\bar C, s = \bar s = 0, S_{\mu} = \partial_{\mu} B,
B_1 + \bar B_1 = 0,  B_2 = \bar B_2 = 0$ in (3.1). Thus,
these values entail upon the expansions
(3.1) to reduce to (4.15) and (4.18). It is obvious that
the horizontality condition
leads to the derivation of the nilpotent (anti-)BRST transformations $s_{(a)b}$
{\it only} for the gauge and (anti-)ghost fields of the theory.}
on the six $(4, 2)$-dimensional supermanifold.

Finally, let us compute the coefficients of the 2-form differentials
$(dx^\mu \wedge dx^\nu)$, constructed by the spacetime variables. The
equality that emerges from l.h.s. and r.h.s., is
$$
\begin{array}{lcl}
\frac{1}{2}\; i e (dx^\mu \wedge dx^\nu)\; \bar \Psi\;
\bigl (\partial_\mu B_\nu^{(g)} - \partial_\nu B_\mu^{(g)} \bigr )\; \Psi
= \frac{1}{2}\; i e (dx^\mu \wedge dx^\nu)\; \bar \psi\;
\bigl (\partial_\mu A_\nu - \partial_\nu A_\mu \bigr )\; \psi.
\end{array}\eqno(4.19)
$$
It is straightforward to check, with the help of
$R_\mu = \partial_\mu C, \bar R_\mu = \partial_\mu \bar C,
S_\mu = \partial_\mu B$, that
$\partial_\mu B_\nu^{(g)} - \partial_\nu B_\mu^{(g)}
 = \partial_\mu A_\nu - \partial_\nu A_\mu$. Thus, the restriction, that
emerges from (4.19), is
\footnote{It is worth emphasizing that the relation (4.20) cannot emerge
from the gauge {\it covariant} version (i.e.
$\tilde D\; \tilde D\; \Psi (x,\theta,\bar\theta)
= D\; D \;\psi (x)$) of the gauge
{\it invariant} restriction (4.1) on the 6D supermanifold.}
$$
\begin{array}{lcl}
 \bar \Psi (x, \theta, \bar\theta) \; \Psi (x, \theta,\bar\theta)
= \bar \psi (x) \; \psi (x) \;\;\Rightarrow \;\; i\; \theta\;
(L_3) + i \;\bar\theta\; (M_3) + i\; \theta \bar\theta\; (N_3) = 0
\end{array}\eqno(4.20)
$$
where the explicit forms of $L_3, M_3$ and $N_3$ are
$$
\begin{array}{lcl}
L_3 = \bar b_2  \psi - \bar \psi \bar b_1\; \qquad M_3 = b_1 \psi
- \bar \psi b_2\; \qquad N_3 = \bar f \psi + \bar \psi f + i \bar
b_2 b_2 - i b_1 \bar b_1.
\end{array}\eqno(4.21)
$$
In the above, the expansions for the fermionic superfields $(\bar \Psi, \Psi)$,
listed in (3.1), have been taken into account for computation of the l.h.s.
$\bar \Psi (x, \theta, \bar\theta) \Psi (x, \theta, \bar\theta)$.

At this juncture, it is worthwhile to mention that the {\it
simple} relationship quoted in (4.20) does not emerge when one
attempts to derive the nilpotent (anti-)BRST symmetry
transformations for all the fields of a given 1-form 4D
non-Abelian gauge theory where there is an interaction between the
1-form non-Abelian gauge field and the Dirac fields. In fact, the
non-Abelian nature of the theory leads to a whole range of
interesting complications when one exploits the restriction (4.1)
on the 6D supermanifold. However, the accurate computation,
ultimately, leads to the derivation of the exact values for the
$b_1, \bar b_1, b_2, \bar b_2, f $ and $\bar f$ present in the
expansions of the superfields $\Psi (x, \theta, \bar\theta)$ and
$\bar \Psi (x, \theta, \bar\theta)$ (cf. (3.1)) for the
non-Abelian gauge theory [18]. This, in turn, leads to the
expansions for the fermionic superfields $\Psi (x, \theta,
\bar\theta)$ and $\bar \Psi (x, \theta, \bar\theta)$ in terms of
the nilpotent (anti-)BRST symmetry transformations for the Dirac
fields of the interacting 1-form non-Abelian gauge theory [18].
Thus, the gauge (i.e. BRST) invariant restriction (4.1) generates
the nilpotent (anti-)BRST transformations for {\it all} the fields
of the interacting Abelian as well as non-Abelian gauge theories
where there is an explicit coupling between the matter fields and
gauge fields.

It is clear that $L_3, M_3$ and $N_3$ of (4.21) should be
separately and independently set equal to zero to maintain the
sanctity of equation (4.1) on the 6D supermanifold. One of the
possible solutions to the conditions: $L_3 = 0, M_3 = 0$ and $N_3
= 0$, is [10] $$
\begin{array}{lcl}
&& b_1 = - e \bar \psi C\; \qquad b_2 = - e C \psi\; \qquad \bar
b_1 = - e \bar C \psi\; \qquad \bar b_2 = - e \bar \psi \bar C
\nonumber\\ && f = - i e\; [\; B + e \bar C C\; ]\; \psi\; \qquad
\bar f = + i e\; \bar \psi\; [\; B + e C \bar C \;].
\end{array} \eqno(4.22)
$$
The solutions, listed in (4.22), form a set of consistent solutions and,
these values,
in some sense,  are very logical
\footnote{To be precise, the solutions in (4.22) are not
the unique set of solutions. This is due to the fact that the signs and
appropriate factors of $i$ and $e$ are {\it not} determined mathematically
in a unique fashion. To obtain the unique set of solutions, the
gauge invariant constraint on the six dimensional supermanifold is:
$\bar \Psi (x, \theta,\bar\theta) (\tilde d
+ i e \tilde A^{(1)}_{(h)}\bigr )
\Psi (x, \theta, \bar\theta) = \bar \psi (x) (d + i e A^{(1)}) \psi (x)$
where $\tilde A^{(1)}_{(h)} = dx^\mu {\cal B}^{(g)}_\mu + d \theta
\bar {\cal F}^{(ac)} + d \bar\theta {\cal F}^{(c)}$. This restriction
on 6D supermanifold has
been exploited in our recent work (see, e.g., [12,13] for details).}.
To elaborate on the above solutions to be a logical one,
let us first focus on $L_3 = 0$ which implies
$\bar b_2 \psi = \bar \psi \bar b_1$. A smart and judicious guess will
be to choose the bosonic components $\bar b_2$ and $\bar b_1$ (of the expansion
in (3.1)) to be proportional to the fermionic fields $\bar \psi$ and
$\psi$, respectively. The latter fields can be made to be bosonic in nature
{\it only} by bringing in the fermionic ($C^2 = \bar C^2 = 0$)
(anti-)ghost fields $(\bar C)C$
of the theory. There is {\it no} other possible choice because
the other fermionic fields ($\psi^2 = 0, \bar \psi^2 = 0$)
of the theory can not do the job. In exactly similar fashion, all the other
choices in (4.22) have been made with an appropriate factors of the
constants $i$ and $e$ thrown in.

It is worthwhile to lay stress, at this stage, that in our earlier
works [10,11] on the consistent extension of the {\it usual}
superfield approach to BRST formalism (endowed with the
horizontality condition alone [1-6]), we exploited an additional
new restriction on the 6D supermanifold by requiring the super
matter current $\tilde J_\mu = \bar \Psi (x,\theta, \bar\theta)
\gamma_\mu \Psi (x,\theta,\bar\theta)$ to be equal to the $U(1)$
gauge invariant and conserved matter current $J_\mu = \bar\psi (x)
\gamma_\mu \psi (x)$. This led exactly to the same kind of
conditions on the component fields of the expansion of $\Psi$ and
$\bar \Psi$, as captured in $L_3 = M_3 = N_3 = 0$. This happened
because of the fact that both the quantities, $\bar \psi
\gamma_\mu \psi$ and $\bar \psi \psi$, are $U(1)$ gauge (and,
therefore, BRST) invariant quantities. The most interesting
feature of our present investigation is the crucial fact that the
condition $\bar \Psi (x,\theta,\bar\theta) \Psi (x, \theta,
\bar\theta) = \bar \psi(x) \psi (x)$ comes out automatically from
the {\it single} restriction (4.1) on the 6D supermanifold which
furnishes the results of the horizontality condition, too. We
would like to lay emphasis on the fact that the condition $\bar
\Psi (x, \theta, \bar\theta) \Psi (x, \theta, \bar\theta) = \bar
\psi (x) \psi (x)$ is superior to the condition $\bar \Psi (x,
\theta, \bar\theta) \gamma_\mu \Psi (x, \theta, \bar\theta) = \bar
\psi (x) \gamma_\mu \psi (x)$ because the former condition is
without the Dirac gamma-matrices whereas the latter condition is
endowed with it. The reason behind the superiority of the former
over the latter is the fact that, so far, we have not been able to
provide a nontrivial six-dimensional representation of the Dirac
gamma-matrices that are present on the l.h.s. of the latter
restriction. It is obvious that the l.h.s. (of the latter
restriction) is defined on the 6D supermanifold.

The insertions of the values of the secondary fields in terms of the basic
fields of the Lagrangian density (2.1), into the super expansion (3.1),
finally, lead to the following expansion of the
superfields in terms of the nilpotent ($s_{(a)b}^2 = 0$)
and anticommuting ($s_b s_{ab} + s_{ab} s_b = 0$) (anti-)BRST
transformations $s_{(a)b}$ of (2.2):
$$
\begin{array}{lcl}
{\cal B}^{(g)}_{\mu}\; (x, \theta, \bar \theta) &=& A_{\mu} (x) +
\;\theta\; (s_{ab} A_{\mu} (x)) + \;\bar \theta\; (s_{b} A_{\mu}
(x)) + \;\theta \;\bar \theta \;(s_{b} s_{ab} A_{\mu} (x))
\nonumber\\ {\cal F}^{(c)}\; (x, \theta, \bar \theta) &=& C (x)
\;+ \; \theta\; (s_{ab} C (x)) \;+ \;\bar \theta\; (s_{b} C (x))
\;+ \;\theta \;\bar \theta \;(s_{b}\; s_{ab} C (x))
 \nonumber\\
\bar {\cal F}^{(ac)}\; (x, \theta, \bar \theta) &=& \bar C (x) \;+
\;\theta\;(s_{ab} \bar C (x)) \;+\bar \theta\; (s_{b} \bar C (x))
\;+\;\theta\;\bar \theta \;(s_{b} \;s_{ab} \bar C (x)) \nonumber\\
\Psi^{(g)}\; (x, \theta, \bar \theta) &=& \psi (x) \;+ \; \theta\;
(s_{ab}  \psi (x)) \;+ \;\bar \theta\; (s_{b} \psi (x)) \;+
\;\theta \;\bar \theta \;(s_{b}\;  s_{ab} \psi (x))
 \nonumber\\
\bar \Psi^{(g)}\; (x, \theta, \bar \theta) &=& \bar \psi (x)
\;+ \;\theta\;(s_{ab} \bar \psi (x)) \;+\bar \theta\; (s_{b} \bar \psi (x))
\;+\;\theta\;\bar \theta \;(s_{b} \; s_{ab} \bar \psi (x)).
\end{array} \eqno(4.23)
$$
The above expressions provide the geometrical interpretations for
(i) the transformations $s_{(a)b}$ (and corresponding generators $Q_{(a)b}$)
as the translational generators along the Grassmannian directions of the
6D supermanifold, (ii) the nilpotency of $s_{(a)b}$ and $Q_{(a)b}$ as
a couple of
successive translations along $\theta$ and $\bar\theta$ directions of
6D supermanifold, and (iii) the anticommutativity properties
of $s_{(a)b}$ and $Q_{(a)b}$ as encoded in the similar
type of relations between translational generators along $\theta$ and
$\bar \theta$ directions.

All the above key properties associated with the (anti-)BRST
transformations for all the basic fields of QED (with Dirac
fields), are encapsulated in the following
$$
\begin{array}{lcl} &&
s_{b}\; \Leftrightarrow\; Q_{b}\; \Leftrightarrow\;
\mbox{Lim}_{\theta \rightarrow 0} {\displaystyle
\frac{\partial}{\partial \bar\theta}} \;\qquad\; s_{ab}\;
\Leftrightarrow\; Q_{ab} \;\Leftrightarrow\;
\mbox{Lim}_{\bar\theta \rightarrow 0} {\displaystyle
\frac{\partial}{\partial \theta}} \nonumber\\ && s_{(a)b}^2 =
0\;\; \Leftrightarrow \;\;Q_{(a)b}^2 = 0\;\; \Leftrightarrow\;
\;\Bigl ({\displaystyle \frac{\partial}{\partial \theta}} \Bigr
)^2 = 0\; \; \Bigl ({\displaystyle \frac{\partial}{\partial
\bar\theta}} \Bigr )^2 = 0 \nonumber\\ && s_b s_{ab} + s_{ab} s_b
= 0\; \Leftrightarrow\; Q_b Q_{ab} + Q_{ab} Q_b = 0
\;\Leftrightarrow \; {\displaystyle \frac{\partial}{\partial
\bar\theta} \frac{\partial}{\partial \theta} +
\frac{\partial}{\partial \theta} \frac{\partial}{\partial \bar
\theta}} = 0.
\end{array}\eqno(4.24)
$$ Thus, all the salient mathematical features of the BRST
symmetries (as well as their generators) have been expressed in
terms of the geometrical objects on the 6D supermanifold.
Furthermore, the derivations of {\it all} the nilpotent
(anti-)BRST symmetry transformations for QED (with Dirac fields)
have been obtained together within the framework of the augmented
superfield formalism in one stroke (cf. (4.1)) and their
geometrical origin and interpretations have been provided.\\

\noindent
{\bf 5 Conclusions}\\

\noindent One of the central results of our present investigation
is the derivation of the nilpotent and anticommuting ($s_b s_{ab}
+ s_{ab} s_b = 0$) (anti-)BRST symmetry transformations $s_{(a)b}$
for the matter (Dirac) fields, the $U(1)$ gauge field and the
(anti-)ghost fields {\it together} from a single restriction (cf.
(4.1)) imposed on the six $(4, 2)$-dimensional supermanifold
(where all the superfields of the theory are defined). This is a
completely new result because, in our earlier works [10-14], the
above nilpotent symmetry transformations have been derived in two
steps by exploiting (i) the horizontality condition, and (ii) its
consistent extensions [10-14], on the 6D supermanifold. It will be
noted, however, that there is an interplay between the above two
restrictions and they are not completely separate and independent.
Thus, for a given $U(1)$ Abelian interacting 4D gauge theory, our
present investigation provides a simpler derivation of the
nilpotent (anti-)BRST transformations for {\it all} the fields of
the theory within the framework of the superfield approach to BRST
formalism.

The new restriction (4.1) on the 6D supermanifold is a gauge {\it
invariant} restriction which leads to the derivation of the
nilpotent (anti-)BRST symmetry transformations for all the fields
(including the matter fields) of QED. Its gauge {\it covariant}
version on the 6D supermanifold does not lead to the derivation of
nilpotent (anti-)BRST symmetry transformations for the matter
(Dirac) fields. It will be noted that the horizontality condition,
on the other hand, is basically a gauge {\it covariant}
restriction on the 6D supermanifold. In fact, the covariant
version of (4.1) leads to the derivation of the exact nilpotent
(anti-)BRST symmetry transformations for the gauge and
(anti-)ghost fields {\it only} which are also the main results of
the restriction due to the horizontality condition on the 6D
supermanifold. Thus, the covariant version of the restriction
(4.1) is equivalent, in some sense, to the restriction due to the
horizontality condition. It is worth emphasizing that the
horizontality condition $\tilde F^{(2)} = F^{(2)}$, reduces to the
gauge {\it invariant} restriction on the 6D supermanifold {\it
only} for the interacting $U(1)$ gauge theory (i.e. QED). This
observation is, however, {\it not} true for the general case of
the interacting  1-form non-Abelian gauge theories.

The importance of the gauge (i.e. BRST) invariant restriction in
(4.1) comes out in its full blaze of glory in the context of
superfield approach to BRST symmetries for the 1-form interacting
non-Abelian 4D gauge theory where there is a coupling between the
1-form non-Abelian gauge field and the Dirac fields [18]. In fact,
it has been shown in this very recent work [18], that the
off-shell nilpotent symmetries for {\it all} the fields (of the
(anti-)BRST invariant Lagrangian density of a given 4D 1-form
interacting non-Abelian gauge theory) can be precisely computed
due to the gauge invariant restriction (4.1) on the 6D
supermanifold. In our present endeavour, there is a great deal of
simplification in the derivation of the nilpotent symmetries for
all the fields of the given 4D 1-form interacting $U(1)$ gauge
theory (i.e. QED). This happens because of its Abelian nature. The
situation is completely different in the case of the superfield
approach to the derivation of the nilpotent symmetry
transformations for the 1-form interacting 4D non-Abelian theory
where the non-Abelian nature of the theory generates interesting
complications (see, e.g. [18]).

The horizontality condition of the usual superfield approach to
BRST formalism has to be generalized so that one could obtain all
the nilpotent symmetry transformations for all the fields of a
given 4D $p$-form (non-)Abelian interacting gauge theory. This is
due to the fact that the results, derived from the application of
the horizontality condition on the 6D supermanifold alone, are
partial in the sense that one obtains only the nilpotent symmetry
transformations for the $p$-form gauge fields and the
corresponding (anti)commuting (anti-)ghost fields of the theory.
The matter fields of the interacting $p$-form gauge theories
remain untouched within the framework of the usual superfield
formalism (with the theoretical arsenal of the horizontality
condition alone). Thus, our present attempt is a step forward in
the direction of the consistent and precise generalization of the
horizontality condition where (i) the nilpotent symmetry
transformations for all the fields (including the matter fields)
of an interacting gauge theory are obtained, and (ii) the
geometrical interpretations for all the properties associated with
the BRST symmetries (and their generators) remain exactly the same
as in the case of the application of the horizontality condition
alone.

As a side remark, it is worthwhile to mention that the present
off-shell nilpotent ($s_{(a)b}^2 = 0$) (anti-)BRST symmetry
transformations $s_{(a)b}$ for the interacting 1-form Abelian
$U(1)$ gauge theory is derived for the specific choice of the
gauge-fixing term (i.e. $-\frac{1}{2} (\partial \cdot A)^2 \equiv
B (\partial \cdot A) + \frac{1}{2} B^2$) in the Feynman gauge. In
this gauge, the ghost fields decouple from the rest of the
physical fields of the theory so that any arbitrary state in the
quantum Hilbert space is a direct product of the physical state
and the ghost states. The subsidiary condition $Q_b |phys> = 0$,
with the conserved and nilpotent BRST charge $Q_b$ on the physical
state (first proposed by Curci and Ferrari [19,20]), plays a
pivotal role in the proof of unitarity of the S-matrix of the
theory by exploiting the so-called ``quartet mechanism'' (see,
e.g., [21] for details). In general, for the non-Abelian gauge
theory, the gauge-fixing term can include the ghost fields and, as
a consequence, there would be an explicit coupling between the
non-Abelian gauge fields and the (anti-)ghost fields. In this
specific case, for the massless as well as massive gauge fields, a
thorough discussion, devoted to the proof of unitarity of the
S-matrix, has been carried out in [20,22,23]. However, for our
present simple case of 1-form Abelian gauge theory in the Feynman
gauge, the physicality criteria ($Q_b |phys> = 0$), the nilpotency
property ($Q_b^2 = 0$) and the conservation of the BRST charge
($\dot Q_b = 0$) are good enough to shed some useful light on the
unitarity of the theory (see, e.g., [21]).

It is interesting to check the validity the idea put forward in
our present investigation, in different contexts (for totally
different kinds of interacting systems). This will enable us to
put our prescription on firmer footings as the gauge invariant
restriction in (4.1) is a general restriction (valid for the
(non-)Abelian gauge theories). In the superfield approach to BRST
formalism, this prescription might be tested for the cases of (i)
the complex scalar fields in interaction with the $U(1)$ gauge
field, (ii) the gravitational theories which are very similar, in
some sense, to the non-Abelian gauge theories (see, e.g., [16] for
details on analogy), and (iii) the 2-form (non-)Abelian gauge
fields and their interactions. Furthermore, it will be a
challenging endeavour to obtain the results of the horizontality
condition and its generalization [12,13] (that lead to
mathematically unique derivations of the nilpotent symmetry
transformations for the matter fields) from a {\it single}
restriction on the 6D supermanifold. The above pointed issues are
some of the promising problems that are presently under
investigation and our results will be reported elsewhere [24].

\end{document}